\documentstyle[preprint,aps]{revtex}
\tighten
\begin{document}
\widetext
\input epsf
\bigskip
\vspace{0.2in}
\thispagestyle{empty}
\begin{flushright}
NYU-TH/99/2/02\\
\today 
\end{flushright}

\vspace{0.2in}

\begin{center}
\bigskip\bigskip
{\large \bf On Field/String Theory Approach to Theta Dependence in 
Large $N$ Yang-Mills Theory}
\vspace{0.3in}      

{Gregory Gabadadze}
\vspace{0.2in}

{\baselineskip=14pt \it 
Department of Physics, New York University, New York, NY 10003, USA} \\
{\it email: gabadadze@physics.nyu.edu } \\
\vspace{0.2in}
\end{center}

\vspace{0.9cm}
\begin{center}
{\bf Abstract}
\end{center} 
\vspace{0.3in}
The theta dependence of the vacuum energy in large $N$ Yang-Mills theory 
has been studied some time ago by Witten using a duality of large $N$ gauge 
theories with string theory compactified on a certain space-time. 
We show that within the field theory context vacuum fluctuations of 
the topological charge give
rise to the vacuum energy consistent with the string theory
computation. Furthermore, we calculate $1/N$ suppressed corrections 
to the string theory result. The reconciliation of the string and 
field theory approaches is based on the fact that the gauge theory instantons
carry zerobrane charge in the corresponding $D$-brane construction
of Yang-Mills theory. Given the formula  for the vacuum energy 
we study certain aspects of stability of the false vacua of the model for 
different realizations of the initial conditions. The vacuum structure appears 
to be different depending on whether $N$ is infinite or, alternatively, large
but finite.

\newpage
\noindent 
{\bf 1. Introduction} 
\vspace{0.2in}

A great deal of information can be learned 
on nonperturbative phenomena in four-dimensional gauge theories
by obtaining these models as a low-energy
realization of certain D-brane configurations \cite {Wittenbranes}, and/or
using  a duality of large $N$ superconformal gauge 
theories and string theory compactified on certain spaces (see Refs. \cite
{Maldacena} and \cite {Polyakov,WittenAdS}). 
This duality, being a powerful technique,
has also been generalized for the case of non-supersymmetric models \cite
{Witten1}. This was applied to study various dynamical issues in large $N$  
pure Yang-Mills theory \cite
{Gross,Ooguri1,Ooguri2,Klebanov,KlebanovTseytlin,Minahan}.  In fact, 
it was used by Witten \cite {WittenTheta} to {\it derive} the expression
for the theta dependence of the vacuum energy  in the large $N$ pure
Yang-Mills (YM) model. The expression for the vacuum energy density associated
with the ${\rm CP}$ odd part of the theory has the form \cite
{WittenTheta}\footnote {It has been conjectured long time ago in \cite
{Witten80}.}: \begin{eqnarray}
{\cal E}_0(\theta) =C ~{\rm min}_k~(\theta + 2\pi k)^2~+ {\cal O} \left 
( {1\over N}  
\right ), 
\label{energy}
\end{eqnarray}
where $C$ is some  constant independent of $N$ and $k$ stands for
an integer number. This expression has a number of 
interesting features which might seem a bit puzzling from the field theory
point of view. Indeed, in the large $N$ limit there are $N^2$ degrees of
freedom in  gluodynamics, thus, naively, one would expect that the vacuum
energy density  in this theory scales as $\sim N^2$.   
However, the leading term in Eq. (\ref {energy}) scales as $\sim 1$. 
As a natural explanation, one could conjecture that there should be 
a colorless massless excitation which saturates the expression for the 
vacuum energy density (\ref {energy}). 
However, pure gluodynamics generates a mass gap and there
are no  physical massless excitations in the model. Thus, the 
origin of Eq. (\ref {energy}) seems to be a conundrum. 

In this work we intend to elucidate the field-theoretic origin of Eq. (\ref
{energy}). In fact, we identify  a very special composite field which
defines  the vacuum energy given in  (\ref {energy}). Moreover, we show
that this composite field does not propagate physical degrees of freedom.
Thus, we clarify the puzzle mentioned above:  The composite field sets the
background energy density but  does not appear in  the spectrum as a physical
state. Using this field we calculate $1/N$ suppressed
corrections to (\ref {energy}).

In section 2, after a brief discussion of the theta dependence in pure
gluodynamics we derive the expression (\ref {energy}) within the field theory
context. Then, we elucidate why the composite field 
defining the background energy (\ref {energy}) does not propagate dynamical
degrees of freedom. In section 3 the same derivation will be given by starting
with QCD  with massive quarks and recovering pure gluodynamics in the
limit when the quark masses go to infinity. Doing so we establish the limits of
applicability of the results of section 2 and, most importantly, we 
calculate the $1/N$ corrections to the vacuum energy density (\ref
{energy}).  In section 4 we discuss the fate of the false vacua emerging in
the theory as a result of the energy dependence given in  ({\ref {energy}).  We
study this issue  for both, the  large but finite  $N$ and infinite $N$ cases. 
Different realizations of the initial conditions for the model  
will be considered. 
In section 5 we discuss the origin of the vacuum energy as it appears in the
string theory and field theory approaches. In fact, we analyze certain 
relations between these two calculations using the arguments of the AdS/CFT
correspondence \cite{Maldacena,Polyakov,WittenAdS}.
Section 6 is devoted to brief discussions and conclusions.  \vspace{0.2in} \\
{\bf 2. Derivation in gluodynamics}
\vspace{0.2in}

The aim of this section is to derive Eq. (\ref {energy}) in pure YM model and,
in fact, to identify the degrees of freedom which are responsible for the theta
dependent vacuum energy density.  

In the quasi-classical  approach the theta
dependence can be  calculated using instantons \cite {BPST}. In a 
simplest approximation  of non-interacting instantons the theta
angle enters the Euclidean space partition function in the following form:
\begin{eqnarray}
\exp \Big (- {8\pi^2\over g^2} \pm i\theta  \Big )\equiv \exp\Big ( -N
{8\pi^2\over \lambda} \pm i\theta \Big )~, 
\label{theta}
\end{eqnarray}
where $g$ stands for the strong coupling
constant. $\lambda$ denotes the 't Hooft's coupling $\lambda\equiv N~g^2$ which
is kept fixed in the large $N$ limit.  The expression above vanishes in the
large $N$ limit, so does the theta dependence in (\ref {theta}). However, this 
conclusion cannot be extrapolated to the infrared region 
of the model. The limitations of the expression (\ref
{theta}) prevent one to do so. Indeed, the quasi-classical approximation is
valid in the limit of small coupling constant (see, for instance, discussions 
in Ref. \cite {Coleman}). Once
quantum corrections are taken into account the coupling constant  $g^2$ in
(\ref {theta}) becomes a  scale dependent quantity. In fact, it will depend 
upon an instanton size $g^2=g^2(\rho)$. For small size instantons the
running coupling is small and the quasi-classical approximation in (\ref
{theta}) holds. However, for large size instantons, i.e. large
couplings, it is not even clear whether the notion of a single instanton is a
legitimate approximation. The overlap  between instantons can be big in this
case and some more complicated field configurations  should be
relevant for the description of physical phenomena \cite {Shuryak}. 
In any event, the expression (\ref {theta}) 
is no longer reliable in  the strong coupling limit.
Thus, the conclusion that the theta dependence goes away in the large $N$ limit
cannot be justified. One way to study the infrared region is to look for some 
appropriate composite colorless excitations for which the notion of  an
asymptotic state can be used. We will start by searching for these
excitations in pure Yang-Mills theory. To proceed, let us recall 
that the topological susceptibility, ${\cal X}$, is a nonzero
number in pure gluodynamics: \begin{eqnarray}
{\cal X} = -i~\int \partial^\mu ~\partial^\nu \langle
0|T~K_{\mu}(x)K_{\nu}(0)|0\rangle~d^4x\neq 0~.  \label{chi}
\end{eqnarray}
Here, $K_\mu $ denotes the Chern-Simons current which is related to the
topological charge density 
as  $Q=\partial^\mu K_\mu = {g^2\over
32\pi^2}F_{\mu\nu}{\tilde F}^{\mu\nu}\equiv F{\tilde F}$, and  the dual
strength-tensor is defined as ${\tilde F}^{\mu\nu}={1\over 2}
\varepsilon^{\mu\nu\alpha\beta}~F_{\alpha\beta}$. All quantities above are
defined in Minkowski space\footnote
{Though in Minkowski space-time  $Q$ does not have the  meaning  of
the topological charge density and, moreover, differs from the Euclidean
definition of the topological charge density by $i$, we formally keep the same
name and letter for simplicity.}. The value of ${\cal X}$ in large $N$ pure
YM theory determines the $\eta'$ meson mass in full QCD with massless quarks. 
This relation, called the Witten-Veneziano formula, can be written as
$m^2_{\eta'} f^2_{\eta'}\propto {\cal X}$, with $f_{\eta'}$ being the $\eta'$
meson decay constant \cite {Witten,Veneziano}\footnote {The numerical
value for ${\cal X}$ can also be found within the framework  of QCD sum
rule \cite {SVZ} calculations \cite {Grunberg}, as well as  in  lattice
studies \cite {latticesusc}. Below, unless otherwise stated,
we will not distinguish between $\cal X $ and its large $N$ limit.
The constant contact term in the definition of $\cal X$ 
will also be omitted for simplicity.}. 

In what follows it will prove convenient   to  introduce  a new variable
by rewriting  the expression for the topological charge density
$Q$ in terms of a four-index (four-form) tensor field  
$H^{\mu\nu\alpha\beta}$:
\begin{eqnarray}
F{\tilde F} = Q = { \varepsilon_{\mu\nu\alpha\beta}H^{\mu\nu\alpha\beta}      
 \over 4!}~, \label{QH}
\end{eqnarray}
where the four-form field $H^{\mu\nu\alpha\beta}$ is the field
strength for the three-form potential $C_{\mu\nu\alpha}$: 
\begin{eqnarray}
H_{\mu\nu\alpha\beta}=\partial_\mu C_{\nu\alpha\beta}-
\partial_\nu C_{\mu\alpha\beta}-\partial_\alpha C_{\nu\mu\beta}-
\partial_\beta C_{\nu\alpha\mu}.  \label{HC} 
\end{eqnarray}
The $C_{\mu\nu\alpha}$ field is defined as a   
composite operator of the gluon fields $A^a_\mu$:
\begin{eqnarray}
C_{\mu\nu\alpha}={1\over 16 \pi^2}(A^a_\mu 
{\overline {\partial}}_\nu
A^a_\alpha-A^a_\nu {\overline {\partial}}_\mu A^a_\alpha-A^a_\alpha 
{\overline {\partial}}_\nu
A^a_\mu+ 2 f_{abc}A^a_\mu A^b_\nu  A^c_\alpha). \label{CA}
\end{eqnarray}
Here, $f_{abc}$ denote the structure constants of the corresponding $SU(N)$
gauge group.  The right-left  derivative in this expression is defined as 
$A{\overline {\partial}}B\equiv A (\partial B)-(\partial A) B $.
Notice, that the $C_{\nu\alpha\beta}$ field is not a gauge invariant
quantity; if the gauge transformation parameter is 
$\Lambda^a$, the three-form field transforms as 
$C_{\nu\alpha\beta}\rightarrow
C_{\nu\alpha\beta}+\partial_\nu \Lambda_{\alpha\beta}-
\partial_\alpha \Lambda_{\nu\beta}-\partial_\beta
\Lambda_{\alpha\nu}$,
where $\Lambda_{\alpha\beta}\propto A_\alpha ^a\partial_\beta
\Lambda^a - A_\beta  ^a\partial_\alpha \Lambda^a$. However, one can  
check that the expression for the field strength $H_{\mu\nu\alpha\beta}$ 
is gauge invariant.

It has been known for some time 
\cite {Luscher} that the $C_{\nu\alpha\beta}$  field 
propagates  long-range correlations if
the topological susceptibility is nonzero in the theory. 
The easiest way to see this is to turn to  
the notion  of the Kogut-Susskind pole \cite {KogutSusskind}.
Let us consider the correlator of the vacuum topological susceptibility at a
nonzero momentum.  
In this case ${\cal X} $ is defined 
as the zero momentum limit of the correlator of two Chern-Simons
currents multiplied by two momenta:
\begin{eqnarray}
{\cal X} =-i~\lim_{q\rightarrow 0}q^\mu q^\nu \int e^{iqx}\langle 0|T
K_\mu (x) K_\nu (0)|0\rangle d^4x.
\label{chinew}
\end{eqnarray}  
Since this expression is nonzero, it must be that the correlator of two 
Chern-Simons  currents develops a pole
as the momentum vanishes, the Kogut-Susskind pole \cite
{KogutSusskind}.

Given that the correlator of two Chern-Simons  currents 
has a pole  and that 
the Chern-Simons current and the three-form 
$C_{\nu\alpha\beta}$  field  are related,
one concludes  that the $C_{\nu\alpha\beta}$ field also has a nonzero
Coulomb  propagator \cite {Luscher}. 
Thus, the $C_{\nu\alpha\beta}$ field behaves as a 
massless collective excitation propagating 
a long-range interaction  \cite {Luscher}. 
These  properties, in the large $N$ limit, 
can  be  summarized in the following 
effective action  for
the $C_{\nu\alpha\beta}$ field:
\begin{eqnarray} 
S_{eff}=-{1\over 2\cdot 4!~{\cal X} }\int
H^2_{\mu\nu\alpha\beta}~d^4x
-{\theta \over 3!} \int_{\partial \Gamma} C_{\nu\alpha\beta}
~dx^\nu \wedge dx^\alpha \wedge dx^\beta +
{\rm High~dim.} 
\label{action}
\end{eqnarray}
The first term in this expression 
yields  the correct Coulomb propagator
for the three-form $C_{\nu\alpha\beta}$ field. The second term is just the 
usual ${\rm CP}$  odd  $\theta$ term of the initial YM action
written as a surface integral at spatial infinity $\partial \Gamma$. 
Notice that higher dimensional  terms are not explicitly 
written in this expression. There might be two types of higher dimensional 
contributions in
(\ref {action}). First of all, there are terms with higher powers of
derivatives of the fields. These terms are suppressed by momenta of the
``massless'' three-form field and do not contribute to the zero momentum vacuum
energy of the system.  In addition, there might be higher dimensional terms 
with no additional derivatives. In the next section we will present some of
these contributions and show that they are suppressed by higher powers of 
$1/N$.

In what follows we are going to study the 
large $N$ effective action  given in Eq. (\ref
{action}) \footnote {The  action (\ref {action})
is not an  effective action in the Wilsonian sense. It is rather
related to the generating functional of one-particle-irreducible
diagrams of the composite field in the large $N$ limit. The 
effective action in Eq. (\ref {action})  is not to be quantized  and loop
diagrams of that action  are not to be taken into
account in calculating  higher order Green's functions. 
The analogous effective action for
the $\rm CP$ even part of the theory was constructed in 
Refs. \cite {Schechter},  \cite{MigdalShifman}, (see also
Ref. \cite {zhitnitski}).}. In particular,  we will calculate
the ground state  energy of the system in the large $N$ limit 
using the effective action (\ref
{action}). In fact, we will derive Eq. (\ref {energy}). 

Before we turn to this
calculation let us mention that Maxwell's equations for a  free four-form
field-strength  $H_{\mu\nu\alpha\beta}$ yield only a constant solution in 
$(3+1)$-dimensional  
space-time \cite {Aurilia}. The reason  is as follows.
A four-form potential has only one independent degree of freedom
in four-dimensional space-time, let us call it $\Sigma$. Then, 
the four Maxwell's equations written in terms of the $\Sigma$ field
require  that this field is independent of the all four space-time 
coordinates, thus the solution can only be a space-time constant.
As a result, the free $H_{\mu\nu\alpha\beta}$ field propagates 
no dynamical degrees of freedom in $(3+1)$-dimensions.
However, this field can be  
responsible for a positive vacuum energy density 
in various  models of Quantum Field Theory (see Ref. \cite {Townsend}).
Thus, studying classical equations of motion for the 
$H_{\mu\nu\alpha\beta}$ field one can  determine  
the value of the ground state energy given by configurations of 
$H_{\mu\nu\alpha\beta}$.   
We are going to solve explicitly the classical equations  of motion
for the effective action (\ref {action}). Then,  
the energy density associated with those solutions will be calculated.

Let us start with the equations of motion. Taking the variation of the 
action (\ref {action}) with respect to the $C_{\nu\alpha\beta}(z)$  
field one gets 
\begin{eqnarray}
\partial^\mu H_{\mu\nu\alpha\beta}(z)=\theta~ {\cal X}~ 
\int_{\partial \Gamma}  \delta^{(4)}(z-x)
~ dx_\nu \wedge dx_\alpha \wedge dx_\beta .
\label{equationmotion}
\end{eqnarray}
This 
equation  can  be solved exactly in four-dimensional space-time \cite
{Aurilia}. The solution is the sum of a particular solution of the 
inhomogeneous equation and a  general solution of the 
corresponding  homogeneous equation:
\begin{eqnarray}
H_{\mu\nu\alpha\beta}(z)=\theta ~{\cal X}~
\int \delta^{(4)}(z-x)
~dx_\mu \wedge  dx_\nu \wedge dx_\alpha \wedge dx_\beta +b~ 
\varepsilon_{\mu\nu\alpha\beta}.
\label{constant}
\end{eqnarray}
The integration constant $b$, if nonzero,  induces an  additional 
CP violation beyond the existed  $\theta$ angle. 
However, periodicity of the $\theta$ angle 
with respect to shifts  by $2\pi \times ({\rm integer})$ allows 
for some nonzero $b$ proportional to  $2\pi {\bf Z}$.
As a result, the general solution to  the equation of motion 
reads as follows:
\begin{eqnarray}
H_{\mu\nu\alpha\beta}=-(\theta +2\pi k)~{\cal X}~
\varepsilon_{\mu\nu\alpha\beta}~.
\label{Htheta}
\end{eqnarray}
Thus, the different vacua are labeled by the integer $k$ and the order
parameter for these vacua in the large $N$ limit 
can be written as: 
\begin{eqnarray}
\langle F{\tilde F} \rangle _k=~(\theta +2\pi k)~{\cal X}.
\label{order}
\end{eqnarray}
As a next step let us 
compute the vacuum energy associated with 
the solution given in Eq. (\ref {Htheta}). The density of the 
energy-momentum tensor for the action (\ref {action})
takes the form
\begin{eqnarray}
\Theta_{\mu\nu}=-{1\over 3!~ {\cal X}    }\left (
H_{\mu\alpha\beta\tau}H_{\nu}^{~\alpha\beta\tau}-{1\over 8}
g_{\mu\nu} H^2_{\rho\alpha\beta\tau}\right ). \label{emt}
\end{eqnarray}
Using the expression (\ref {Htheta})
one calculates the corresponding 
energy density\footnote {
Notice that the total YM energy density should contain some  negative
constant related to the nonzero value of the gluon condensate \cite {SVZ}.
This constant is subtracted from the expression for the energy discussed in
this work. The energy density (\ref {newenergy}) is normalized as 
${\cal E}_0(\theta=0)=0$, and 
for $k=0$ was discussed in \cite {membrane}.}  ${\cal E}_k$ 
\begin{eqnarray}
{\cal E}_k(\theta)={1\over 2}(\theta+2\pi k)^2~ {\cal X}. 
\label{newenergy}
\end{eqnarray}
Since the $H_{\mu\nu\alpha\beta}$ field does not propagate 
dynamical degrees of freedom the expression above is the 
total energy density of the system given by the action  (\ref {action}) 
\footnote {One  might wonder
whether the same result is obtained if one treats
$\theta$ not as a constant multiplying  $Q$ in the Lagrangian,
but as the phase that the states acquire under a topologically
non-trivial gauge transformations. 
In this case the arbitrary integration constant 
in Eq. (\ref {constant}) has to be chosen in such a way
which would guarantee a  
proper $\theta$ dependence of the VEV of the topological charge density.
This would leave the results of our discussion without modifications.}.   

Before we go further let us stop here to discuss some of the consequences
of Eq. (\ref {newenergy}). First of all, let us notice that 
the result (\ref {newenergy}), as well as Eq. (\ref {energy}),  
is only valid  in the 
limit of infinite $N$. In the next section we  will calculate 
subleading order corrections to  Eqs. (\ref {energy},\ref {newenergy})
and argue that these expressions can also  be used as a good approximation
for large but finite $N$. The constant $C$ emerging 
in (\ref {energy}) is related to the topological susceptibility as follows:
$$ {\cal X}|_{N\rightarrow \infty}=2 C.$$
Thus, the vacuum energy (\ref {energy},\ref {newenergy}) 
is defined by vacuum fluctuations of
the topological charge measured by $\cal X$.  

The crucial feature of (\ref {newenergy}) is that it defines an infinite number
of vacua. The true vacuum is obtained by minimizing  (\ref
{newenergy}) with respect to $k$ as in (\ref {energy}):
\begin{eqnarray}
{\cal E}_0(\theta)={1\over 2}~ {\cal X}~{\rm min}_k~
(\theta+2\pi k)^2~.
\nonumber
\end{eqnarray}
This expression is periodic with respect to shifts of $\theta$ by
$2\pi {\bf Z}$ and is also a smooth function of $\theta$ except
for $\theta=\pi$ \cite {WittenTheta} (see also discussions below).
Thus, there are an
infinite number of the false vacua in the theory \cite {WittenTheta}. 
The fate of these states will be discussed in section 4. 
\vspace{0.2in} \\
{\bf 3. Large $N$ QCD calculation}
\vspace{0.1in}  

In this section we consider full QCD with three quark flavors. We are going to
write down a low-energy effective Lagrangian for this case and then gradually
decouple quarks by taking the quark masses to infinity. The resulting effective
Lagrangian should be giving the energy density for pure Yang-Mills theory. 

In the  large $N$ expansion the effective Lagrangian of 
QCD with three flavors takes the form \cite
{VenezianoDi,Arnowitt,Trachern,Witten80}: \begin{eqnarray}
{\cal L}(U,U^*,Q)={\cal L}_0(U,U^*)+
{1\over 2 }~ i~ Q (x)~ {\rm Tr}~ \Big ( {\rm ln}~U-{\rm ln}~U^*\Big )+ 
\nonumber  \\
{1\over 2~{\cal X}}~ Q^2(x) +\theta~ Q(x) +{B \over 2 \sqrt{2} }~{\rm Tr}
~(MU+M^*U^*)+\dots ~, 
\label{effective}
\end{eqnarray}
where $U$  denotes the flavor group matrix of pseudoscalar mesons,
${\cal L}_0$ denotes the part of the Lagrangian which
contains the meson fields only \cite {VenezianoDi,Arnowitt,Trachern,Witten80},
$B$ is some  constant related to the quark condensate,
and $M$ stands for the meson mass matrix (for recent discussions of
the effective chiral Lagrangian approach see Ref. \cite {Pich}). Higher order
terms in (\ref {effective}) are suppressed by quadratic and higher powers of
$1/N$.  In order to study vacuum properties, we concentrate  
on the low-momentum approximation. 
The Lagrangian presented above can be used to solve the $U(1)$ problem \cite
{Witten,Veneziano}. Indeed, the field $Q$ enters the Lagrangian in a quadratic
approximation and can be integrated out. As a result, the flavor singlet
meson, the $\eta'$, gets an additional contribution into its mass term.
This leads to the Witten-Veneziano relation and the solution of the $U(1)$
problem without instantons \cite {Witten,Veneziano}. In the present case we
would like to follow an opposite way. Namely, we are going to make quarks
very heavy and integrate them out keeping the field $Q$ in the Lagrangian.
In the limit $m_q\rightarrow \infty$ one finds that $M\rightarrow
\infty$. Thus, the low-energy effective Lagrangian which is left after the
mesons are integrated out will take the form:  
\begin{eqnarray} 
{\cal L}_{\rm eff} (Q)=
{ 1\over 2~{\cal X} }~ Q^2(x) +\theta~ Q(x) +
{\cal O}~ \Big ( { \Lambda^2_{\rm QCD} \over M^2 },~~{\partial^2 Q^2\over
\Lambda^6_{\rm QCD}},~~{1\over N^2} \Big )~.  \label{effectiveYM}
\end{eqnarray}
Rewriting the field $Q$ in terms of the ``massless'' 
tensor $C_{\alpha\beta\gamma}$ as in
the previous section,  one finds that   
the expression (\ref {effectiveYM}) is nothing but the Lagrangian presented in
(\ref {action}). Thus, the higher order terms neglected in 
(\ref {action}) which could contribute to the vacuum energy at zero
momenta would correspond to higher corrections in $1/N$. In fact, the
subleading  corrections to  the effective Lagrangian (\ref {effective}) can
also be found \cite {Pettorino}. These terms are proportional (with the
corresponding dimensionful coefficients) to the following expressions: 
\begin{eqnarray}
{ {\rm const.} \over N^2 }~ Q^2~{\rm Tr}~ 
(\partial_\mu U~\partial_\mu U^*),
~~~~~~~~~~~{ {\rm const.}\over N^2 }~ Q^4~. 
\label{corrections}
\end{eqnarray}
The terms listed above are suppressed in the effective Lagrangian 
by the factor $1/N^2$. 
As a next step, we can include the terms (\ref {corrections}) into the full
effective Lagrangian and then integrate the heavy meson fields out. The net
result of this procedure is that the terms proportional to $Q^4$ appear in the
effective Lagrangian for pure YM theory. This, in its turn,
modifies the equation of motion for the single component 
of $H_{\mu\nu\alpha\beta}$ considered in the previous section.
 Performing the calculation of the vacuum
energy in the same manner as in section 2  we find the following result
for the energy density:
\begin{eqnarray} 
{\cal E}_k(\theta)={1\over 2}~ {\cal X} ~\Big ( 
\theta+2\pi k\Big )^2+ {{\rm const.}\over N^2}~ {\cal X}~ 
\Big (  \theta+2\pi k\Big )^4~+~{\cal O} \Big ( {1\over N^3} \Big ). 
\label{1overN}
\end{eqnarray}
In this expression the arbitrary constant emerges as a result of integration
of the equation  of motion\footnote {It is not clear, however, what is the
numerical value of this constant, and whether in fact it is nonzero.}.  
Notice that the topological susceptibility in the 
expression above is also defined in the corresponding order in the large 
$N$ expansion: ${\cal X}=2 C+{\cal X}_{1}/N+{\cal X}_2/N^2$.
Thus, the expressions (\ref {newenergy},\ref {1overN}) could in
principle  give a reasonable approximation for big enough but otherwise {\it
finite} $N$. 
The true vacuum energy density, ${\cal E}_0(\theta)$, can be obtained by
minimizing the  expression (\ref {1overN}) with respect to $k$ as in (\ref
{energy}). Then, ${\cal E}_0(\theta)$ satisfies the relation 
$\partial^2_{\theta}
{\cal E}_0(\theta)~|_{\theta=0} ={\cal X} $, no matter what 
is the value of the
arbitrary constant in (\ref {1overN}).  
\vspace{0.2in} \\
{\bf 4. Dynamics of the false vacua}
\vspace{0.1in} 

In this section we will discuss the dynamics of the false vacua present 
in the theory. In accordance  with (\ref {newenergy},\ref {1overN})            there are an infinite number of vacua for any given value of 
the theta angle \cite
{WittenTheta}.  Clearly, not all of these are degenerate. 
The true vacuum state is defined by minimizing the expressions  
(\ref {newenergy},\ref {1overN}) 
with respect to $k$. All the other states are false
vacua with greater values of the energy density. There is a potential barrier
that separates a given false vacuum state from the true one. Thus, a false
vacuum can in general  decay into the true state through the process of 
bubble nucleation \cite {bubble}\footnote {This decay can go through the
Euclidean ``bounce'' solution \cite {bounce}. Though the  existence of
the bounce for this case is not easy to understand  within the field theory
context, nevertheless, one could  be motivated by the brane construction where
this object appears as a sixbrane bubble wrapped on a certain 
space \cite {WittenTheta}.}. The decay rates for these vacua  were
evaluated recently in  Ref. \cite {Shifman}.  In this section
we analyze  the fate of the false vacua
for different realizations of the initial conditions in which the system is
placed. For the sake of simplicity  we will be discussing transitions between
the vacuum states labeled by $k'$ and $k$ for different  values 
of these integers. 
The first two cases considered in this section were studied in Refs. \cite
{WittenTheta} and \cite {Shifman}, we include them here for 
completeness.   
\vspace{0.2in} \\ 
{\it 4.1. The false vacua with
$k'\sim 1$}  \vspace{0.1in} 

In this subsection we consider the system which in its initial state 
exists in a 
false vacuum with $k'$ of order  $\sim 1$. Let us start with the case 
when $N$ is a large but finite number so that the formula (\ref
{newenergy}) (or (\ref {1overN})) is still a good approximation. Since there
exists the true vacuum state with less energy, the false vacuum can ``decay''
into  the true one 
via the bubble nucleation process. That is to say, there is a finite 
probability to form
a bubble with the true vacuum state inside.  The shell of the bubble 
is a  domain wall which separates the false state from the true one.
The dynamical question we discuss here is whether it is 
favorable energetically to create and expand such a bubble. 
Let us study the energy balance for the case at hand. While creating the shell
of the bubble one looses the amount of energy equal to  the surface area
of the bubble multiplied by the tension of the shell. On the other hand, the
true vacuum state is created inside the bubble, thus, one gains  the amount of
energy equal to the difference between the energies of the false and true
states.  The energy balance between these two effects defines whether
the bubble can be formed, and, whether the whole false vacuum 
can transform into
the true one by expanding  this bubble to infinity. 
Let us start with the volume energy density. 
The amount of the  energy density which is gained
by creating the bubble is\footnote{In this subsection we 
assume that $\theta
\neq \pm \pi$. The case  $\theta=\pi$ will be considered below.}:
\begin{eqnarray}
\Delta {\cal E} ={1\over 2}~ {\cal X}~ 
\Big [ (\theta +2 \pi)^2-\theta^2 \Big ]=2
\pi~ {\cal X}~ (\theta+\pi).
\nonumber
\end{eqnarray}  
Thus, $\Delta {\cal E}$    scales as $\sim 1$ in the large $N$ limit as
long as the volume of the bubble is finite. Let us now
turn to the surface energy which is lost. This energy
is defined as: 
\begin{eqnarray}
{E}_s= T_D~\times ({\rm surface ~~area})~.
\label{surface}
\end{eqnarray}
The tension of the wall between the adjacent vacua, $T_D$, should scale as
$T_D\sim N$ in the large $N$ limit \cite {WittenTheta}. Hence, the surface
energy will also scale as $\sim N$. Thus, the process of creation of 
a finite volume bubble in the  large $N$ limit is not energetically favorable.
Indeed, 
the amount of energy which is lost while creating the shell
is bigger than the amount which is gained. 
In terms of the false vacuum decay width 
this means that the width of this process is suppressed in the 
large $N$ limit  
\cite {Shifman}: \begin{eqnarray}
{\Gamma\over {\rm Volume} }~\propto ~ \exp \Big (- a N^4 \Big ),
\label{width}
\end{eqnarray}
where $a$ stands for some positive constant \cite {Shifman}.
Thus, one concludes that in the limit $N\rightarrow\infty$  the false vacua
with $k'\sim 1$ are stable \cite {WittenTheta,Shifman}.
\vspace{0.2in} \\
{\it 4.2. The false vacua with $k'\sim N$}
\vspace{0.1in}

Here we study the fate of the false vacua with $k'\sim N$. We discuss 
a possibility that these vacua  can decay into a state $k$
with $k'-k\sim N$ and $k'+k\sim N$.  As in the previous subsection, we are
going to study the energy balance for the bubble nucleation process. The amount
of the volume energy density which is gained by creating  such a bubble  
in the large $N$ limit scales as follows:
\begin{eqnarray}
\Delta {\cal E}~\propto ~ {\cal X}~ N^2.
\nonumber
\end{eqnarray}
Thus, the volume energy which is gained increases as $\sim N^2$. Let us now
turn to the surface energy which is lost while nucleating a bubble. This is
defined as ${E}'_s=T'_D\times ({\rm surface~~area})$, where $T'_D$
denotes the  tension of the domain wall interpolating between the vacua
labeled by $k'$ and $k$. Since $k'-k\sim N$ these vacua are not neighboring
ones. Thus, in general, there is no reason to expect that the tension of the
wall interpolating between these vacua scales as $\sim N$. 
$T'_D$ might scale
as $\sim N^2$ at most (as the energy of a generic
configuration in a model  with $N^2$ degrees of freedom). 
However, even in the case when $T'_D\sim N^2$ 
the volume energy which is gained is at least of the same order 
as the  surface energy which is lost. Hence, 
it is energetically favorable to increase the radius of such a bubble  
(since the volume energy scales as a cubic power of the radius  while the
surface energy scales only as a quadratic power of the effective size). 
Thus, the bubble nucleation process will not be suppressed and the false
vacua with $k'\sim N$ will eventually decay into the true ground state. 
Note, that the state $k'=N$ can as well decay into the neighboring vacuum
$k=N-1$  which subsequently is allowed to turn into the ground state. 
\vspace{0.2in} \\
{\it 4.3. Parallel domain walls}
\vspace{0.1in} 

In this subsection we consider the special case when all the vacua are
present simultaneously in the initial state of the model. This can be
achieved, for instance, by placing in space an infinite number of parallel
domain walls  separating  different vacua from each other. It is
rather convenient to picture these walls as parallel planes. Each vacuum state
is sandwiched between the corresponding two domain walls (two planes)
separating this state from the neighboring vacua. Each domain is
labeled by $k$ and in accordance with (\ref {newenergy},\ref {1overN}) is
characterized by the corresponding value of the vacuum energy. Furthermore,
the order parameter $\langle F{\tilde F}\rangle $ takes different values in
these vacua in accordance with (\ref {order}).  Let us turn to the true vacuum
state. For simplicity we assume  that this state is given by $k=0$ (which
corresponds to $|\theta|$  being less than $\pi$). The corresponding vacuum
energy is the lowest one.  Consider the two states which are adjacent to the
true vacuum.  These states have the energy density bigger than that of
the true vacuum.  Thus, there is a constant pressure acting on the domain walls
separating the true vacuum from the adjacent false ones. This pressure will
tend to expand the domain of the true vacuum. In fact, for large but finite 
$N$,  the pressure will indeed expand the spatial region of the true vacuum 
by  moving apart the centers of the domain walls sandwiching this 
state.  The very same effect will be happening  between any two adjacent
vacua. Indeed, let us calculate the jump of the energy density between the two
vacua labeled by $k'$ and $k$:  
\begin{eqnarray} 
\Delta {\cal  E}_{k'k}=2\pi~ {\cal X}~ (k'-k) ~\Big (
\theta+\pi(k'+k) \Big )~. 
\label{deltae} 
\end{eqnarray}
As far as $N$ is large but finite, the walls will start to accelerate. 
Farther the wall is located from the true vacuum (i.e. larger the sum
$k'+k$),  bigger the initial acceleration of the wall is going to be;
i.e., the walls will start to move apart from each other with the following
initial acceleration: 
\begin{eqnarray}
a_{k'k}\propto \Lambda_{\rm  YM }~ {(k'-k)~ [\theta +\pi (k'+k)] \over N}~.
\label{acceleration}
\end{eqnarray}
For finite $N$ all the walls will be moving to spatial infinity and the 
whole space will eventually be filled with the true vacuum state. 
On the other hand,
when $N\rightarrow \infty $ the picture is a bit different.
There are a number of interesting cases to consider:  

First of all let us
set $k'-k =1$ and $k'~,k \sim 1$. Then, in the limit  $N\rightarrow \infty $
the acceleration $a_{k'k}\rightarrow 0$.
Thus, the neighboring walls  stand still if they had no initial velocity.
The physical reason of this behavior is as follows\footnote{I am grateful
to Gia Dvali who pointed this out to me.}: Though there is a constant pressure
of order $\sim 1$ acting on the wall, nevertheless, 
the wall  cannot be moved because
the mass per unit surface area of the wall
tends to infinity in the limit $N\rightarrow \infty$. 

The second interesting case would be  when the
constant pressure produced by the energy jump between some neighboring 
vacua is
of order $\sim N$. In this case it will  be possible to accelerate
these walls up to the speed of light and send them to spatial infinity.
Indeed, if $k'-k =1$ but $k'+k \sim N$, then the wall
between these two vacua starts moving with a non-vanishing acceleration
which scales as follows:
\begin{eqnarray} 
a_{k'k}\propto \Lambda_{\rm YM} {\pi (k'+k) \over N}\sim
{\cal O} (1).  \label{nonumber}
\end{eqnarray}
Thus, these walls will eventually be approaching spatial infinity with a speed
of light even in the limit of infinite $N$. 

In addition to  the effects emphasized above 
there might also be decays of the false
vacua happening in each particular domain. As we discussed in the
previous subsections, for large but finite $N$ all the false vacua will be
nucleating bubbles with energetically favorable phases inside and
expanding these bubbles to infinity. Thus, for large but finite $N$, there are
two effects which  eliminate the false vacua: The moving walls
are sweeping these states to infinity, and, in addition, these vacua are
decaying via bubble nucleation processes. 

What happens for an infinite $N$?
As we learned above there are an infinite number of domains which will stay
stable in that limit  and the corresponding false vacua would not decay because
of the exponential suppression. Thus, there are an infinite
number of inequivalent spatial regions which are separated 
by domain walls. 
Consider one of the regions sandwiched between two domain walls.
The three-form field $C_{\mu\nu\alpha}$ will couple to the walls
and the large $N$ effective action for this case will look as follows:
\begin{eqnarray}
{\tilde S} =S_{\rm eff}+\sum_{i=k,~k+1}~\mu_i~\int_{{\cal W}_i} 
C_{\mu\nu\alpha}~dx^{\mu} \wedge dx^{\nu} \wedge dx^{\alpha}, 
\label{tildeS}
\end{eqnarray}
where $S_{\rm eff}$ is defined in (\ref {action}), $\mu_i$ stands 
for the coupling of the three-form potential to a corresponding domain wall;
${\cal W}_i$ denotes the worldvolume of the wall. In this case the domain
wall can be regarded as a source of the corresponding three-form potential.
This is reminiscent to what happens in the large $N$  supersymmetric YM model 
\cite {GiaZuraMe}.      
\vspace{0.2in} \\
{\it 4.4. Domain walls at  $\theta=\pi$}
\vspace{0.1in}  

If $\theta=\pi$, the initial classical Lagrangian is ${\rm CP}$ invariant.
Indeed, under ${\rm CP}$ transformations $\theta=\pi$ goes into $-\pi$. 
Since $\pi$ and $-\pi$ angles are equivalent,  ${\rm CP}$ is a symmetry of
the Lagrangian.  However, in accordance with (\ref {order}), 
this symmetry is
spontaneously broken by the vacuum of the theory. Thus, one finds the
following two degenerate true vacua:
\begin{eqnarray}
{\cal E}_{k=0}={\cal E}_{k=-1}={1\over 2}~ {\cal X}~ \pi^2~.
\label{degenerate}
\end{eqnarray}
These two vacua are labeled by the order parameter
(\ref {order}).  In the $k=0$ state  $\langle F{\tilde F}\rangle =
\pi{\cal X}$ and in the $k=-1$ state  $\langle F{\tilde F}\rangle =
-\pi{\cal X}$.  As a result of the spontaneous breaking of a discrete symmetry
there should be a domain wall separating these two vacua.  Let us consider
the case discussed in the previous section.  Namely, let us choose the initial
condition of the system as a state where all the possible vacua are
simultaneously realized in space. That is, there are an infinite number of
domain walls (parallel planes) dividing space into an infinite number of
domains with different values of the vacuum energy density labeled
by $k$.  As we mentioned
above, only two of these domains have equal minimal energy density 
given in (\ref {degenerate}). 
The domain wall separating these two vacua, as we will see below, is somewhat
special. In accordance with the discussions in the
previous subsection, for large  but finite $N$, all the walls merging with
the false vacua will  tend to rush to spatial infinity. 
The final stable state of
the model can be described as a space separated into two parts by a single
domain wall. To the left (right) of the wall one  discovers the phase with
$k=-1$ with the corresponding order parameter $\langle F{\tilde F}\rangle =
-\pi{\cal X}$ , and, to the right (left) of the wall one finds 
the state with $k=0$ and $\langle F{\tilde F}\rangle =
\pi{\cal X}$.  Vacuum energies of these two states are degenerate.  

In the case of infinite $N$ the picture is slightly different.  As elucidated
in the previous subsection, there will be an infinite number of stable vacua. 
The domain wall separating  the two true vacua can be regarded in this case as 
the fixed plane under ${\bf Z}_2$ transformations of the coordinate
transverse to the plane. 
The  three-form field $C_{\mu\nu\alpha}$ will be able to 
couple to this wall in a manner discussed in the previous subsection.
The picture, being interpreted in the string theory approach, looks much
like the one with an orientifold plane separating  mirror $D$-branes on its
both sides \cite {orientifold}.
\vspace{0.2in} \\
{\bf 5. A Comment on String Theory vs. Field Theory Approach}
\vspace{0.1in}  

We found in section 2 that the theta dependent vacuum energy (\ref {energy})
is related to the correlator (\ref {chi}) measuring the vacuum fluctuations of
the topological charge. The question which arises here is whether this can be
seen from the original string theory computation \cite
{WittenTheta}. We are going to discuss below  how the 
string theory calculation indicates that the vacuum energy
(\ref {energy}) should indeed be related to the vacuum fluctuations of the
topological charge. In fact, we argue that this is related to the instantons
carrying $D0$-brane charge in the Type IIA fourbrane construction of the
four-dimensional YM model.

To begin with let us recall how the theta dependent vacuum energy  appears in
the brane construction of the four-dimensional YM model \cite {WittenTheta}.
One starts with Type IIA superstring theory on 
${\cal M}\equiv {\bf R}^4\times {\bf S}^1\times {\bf R}^5$, 
with $N$ coincident $D4$-branes \cite {Witten1}.
The $D4$-brane worldvolume is assumed to be $ {\bf R}^4 \times {\bf S}^1$
and the fermion boundary conditions on ${\bf S}^1$
are chosen in such a way that the low-energy theory on the worldvolume is
pure non-supersymmetric $U(N)$ YM theory \cite {Witten1}. 
In the dual description, 
the large $N$ limit of the $SU(N)$ part of this theory can be studied by
string  theory on a certain background \cite
{Maldacena,Polyakov,WittenAdS,Witten1}. It was shown in Ref. \cite
{WittenTheta} that the theta dependent vacuum energy (\ref {energy}) 
arises in the dual string description due to  the  $U(1)$ gauge field
$B_M,~M=1,..,5$. To find out what this corresponds to in the original gauge
theory language  recall that this $U(1)$ field is nothing but the
Ramond-Ramond (RR) one-form of Type IIA theory. Furthermore, once the gauge
theory is realized in the  Type IIA fourbrane construction, the 
Wess-Zumino-Witten (WZW) term present in the worldvolume effective action
defines the correspondence between the gauge theory operators on one
side and the string theory Ramond-Ramond fields on the other side. 
In the case at
hand the worldvolume WZW term  looks as follows:
\begin{eqnarray}
{S}_{\rm WZW}={1\over 8\pi^2}~\int_{\cal V}~B\wedge {\rm Tr}F\wedge F,
\label{WZW}
\end{eqnarray}
where ${\cal V}$ denotes the worldvolume of a  wrapped fourbrane,
${\cal V}\equiv {\bf R}^4\times {\bf S}^1$. In accordance with 
the general principles of the large $N$ AdS/CFT correspondence \cite
{Maldacena,Polyakov,WittenAdS} the
classical action for the RR one-form  on the string theory
side defines the YM correlation 
functions of the composite operator $F{\tilde F}$
(since this is the operator which couples to the corresponding RR field in
(\ref {WZW})). Moreover, the two-point
correlator of  the topological charge density  $F{\tilde F}$, which coincides
with  ${\cal X}$ in (\ref {chi}) up to some contact term \cite {Witten}, was
recently calculated using  the AdS/CFT correspondence and the supergravity
analyses for the RR one-form $B_M$ \cite {HashimotoOz}. Thus, it is not
surprising that the theta dependent vacuum energy which is defined by the RR
one-form in the string theory calculation is related to the nonzero value of
the topological susceptibility (\ref {chi}) in the gauge theory studies.    
The physical reason for this correspondence, as we have mentioned above,
is the special property of the gauge theory instantons in 
the fourbrane construction. 
Indeed, in accordance with (\ref {WZW}) the RR one-form
couples to the topological charge density $F{\tilde F}$, on the other hand 
the RR one-form couples by definition to $D0$-branes. Thus, the gauge theory
instantons in this case carry zerobrane charge. This is  the physical reason
for the correspondence discussed above.     
\vspace{0.2in} \\
{\bf 6. Conclusions}
\vspace{0.1in}

Let us summarize briefly the results of this work.
The expression for the theta dependent vacuum energy 
derived in Ref. \cite {WittenTheta} in a string theory calculation was
obtained in large $N$ pure YM theory. This energy is defined by
vacuum fluctuations of the topological charge. 
The field-theoretic  approach  allows one to  clarify the puzzle
why the theta dependent vacuum energy is of order $\sim 1$ 
in the large $N$ limit.  
The field which saturates the expression for the 
energy has very specific properties, 
it produces the  background energy
density but does not propagate any dynamical degrees of freedom
in the large $N$ limit.  
Within the field theory framework subleading 
corrections to the large $N$ result (\ref {energy}) were 
calculated  (see Eq. (\ref {1overN})). 

Given the expression for the theta dependent vacuum energy
we studied  the issue of stability of the false vacua for different
realization of the initial conditions of the system.  The large but finite $N$
and infinite $N$ cases were considered. The structure of the vacuum state in
these two cases is rather  different. 

The string theory
calculation provides important clues on what the theta dependent vacuum
energy should be related to within the field theory approach. We discussed
this issue arguing that the reconciliation of the field theory 
result with the original string theory calculation is based 
on the fact that instantons carry  $D0$-brane charge
in the $D4$-brane construction of pure YM model.   
\vspace{0.1in} 
\vspace{0.2in} \\
{\bf Acknowledgments}
\vspace{0.1in}

The author is grateful to Gia Dvali for reading the 
manuscript and for useful discussions and suggestions. The
work was supported in part by the grant NSF PHY-94-23002.


\vspace{0.1in} 

\begin{references}

\bibitem{Wittenbranes} For a recent review, see: \\
A. Giveon, D. Kutasov, ``Brane Dynamics and Gauge Theory'', hep-th/9802067.

\bibitem{Maldacena} J. Maldacena, ``The Large $N$ limit of Superconformal
Field Theories and Supergravity'', Adv. Theor. Math. Phys. {\bf 2}
(1998) 231;  hep-th/9711200. 

\bibitem{Polyakov} S.S. Gubster, I.R. Klebanov, A.M. Polyakov, ``Gauge Theory
Correlators from Noncritical String Theory'', Phys. Lett. {\bf B428}
(1998) 105; hep-th/9802109. 

\bibitem{WittenAdS}  E. Witten, ``Anti-de Sitter Space and Holography'',
Adv. Theor. Math. Phys. {\bf 2} (1998) 253;
hep-th/9802150. 

\bibitem{Witten1} E. Witten, ``Anti-de Sitter Space, Thermal
Phase Transitions, and Confinement in Gauge Theories'', 
Adv. Theor. Math. Phys. {\bf 2} (1998) 505; 
hep-th/9803131.

\bibitem {Gross} D. J. Gross, H. Ooguri, ``Aspects of Large $N$ Gauge Theory
Dynamics as Seen by String Theory'', Phys. Rev. {\bf D58} 
(1998) 106002; hep-th/9805129.

\bibitem{Ooguri1} C. Csaki, H. Ooguri, Y. Oz, J. Terning, ``Glueball Mass
Spectrum from  Supergravity'', hep-th/9806021;
R. de Mello Koch, A. Jevicki, M. Mihailescu, J.P. Nunes,
``Evaluation of Glueball Masses from Supergravity'', Phys. Rev. {\bf D58}
(1998) 105009; hep-th/9806125; 
M. Zyskin, ``A Note on
the Glueball Mass Spectrum'', Phys. Lett. {\bf B439} (1998) 373; 
hep-th/9806128; 

\bibitem{Ooguri2} H. Ooguri, H. Robins, J. Tannenhauser, ``Glueballs and their
Kaluza-Klein  Cousins'', Phys. Lett. {\bf B437} (1998) 77.

\bibitem{Klebanov} I.R. Klebanov, ``From Threebranes to Large $N$ Gauge
Theories'', hep-th/9901018.

\bibitem{KlebanovTseytlin} I.R. Klebanov, A.A. Tseytlin, ``$D$-branes and Dual
Gauge Theories in Type $0$ Strings'', hep-th/9811035.

\bibitem{Minahan} J. Minahan, ``Glueball Mass Spectra and Other Issues for
Supergravity Duals of QCD Models'', hep-th/9811156.

 \bibitem{WittenTheta} E. Witten, ``Theta dependence in the Large
$N$ Limit of Four-Dimensional Gauge Theories'', 
Phys. Rev. Lett. {\bf 81} (1998) 2862; hep-th/9807109. 

\bibitem{Witten80} E.Witten, ``Large $N$ Chiral Dynamics'', Ann. of Phys. 
{\bf 128} (1980) 363.  

\bibitem{BPST} A.M. Belavin, A.M. Polyakov, A.S. Schwartz,
Yu.S. Tyupkin,
``Pseudoparticle Solutions of the Yang-Mills Equations'', 
Phys. Lett. {\bf 59B} (1975) 85. 

\bibitem{Coleman}   S. Coleman, ``Aspects of Symmetry'', Cambridge
University Press, 1985.

\bibitem{Shuryak} E.V. Shuryak, ``The Role of Instantons in Quantum 
Chromodynamics. 1. Physical Vacuum'',  
Nucl.Phys. {\bf B203} (1982) 93; \\
D.I. Diakonov, V.Y. Petrov, ``Instanton Based Vacuum from
Feynman Variational Principle'',
Nucl.Phys. {\bf B245} (1984) 259; \\
T. DeGrand, A. Hazenfratz, T.G.  Kovacs,
``Topological Structure in the $SU(2)$ Vacuum'',
Nucl. Phys. {\bf B505} (1997) 417; hep-lat/9705009.

\bibitem{Witten} E. Witten, `` Current Algebra Theorems for the 
U(1) ``Goldstone Boson'' '',  Nucl. Phys. {\bf B156} (1979) 269.

\bibitem{Veneziano} G. Veneziano, ``U(1) Without Instantons'', 
Nucl. Phys. {\bf B159} (1979) 213.

\bibitem{SVZ} M.A. Shifman, A.I. Vainshtein, V.I. Zakharov, 
``QCD and Resonance Physics. Theoretical Foundations'', Nucl. Phys.
{\bf B147} (1979) 385. 

\bibitem{Grunberg} G. Grunberg, ``QCD Sum Rule Determination of 
the Topological Susceptibility'', Phys. Rev {\bf D30 } (1984) 1570. 

\bibitem{latticesusc} B. Alles, M. D'Elia, A. Di Giacomo, 
``Topological  Susceptibility  at  Zero  and  Finite $T$  in $SU(3)$
Yang-Mills Theory'', Nucl. Phys. {\bf B494} (1997) 281;  
hep-lat/9605013. 
  

\bibitem{Luscher} M. L\"uscher, ``The Secret Long Range Force in
Quantum Field Theories With Instantons'', Phys. Lett. {\bf 78B} (1978) 
465.

\bibitem{KogutSusskind} J. Kogut, L. Susskind, ``How to Solve the 
$\eta\rightarrow  3\pi$ Problem by Seizing the Vacuum '',  
Phys. Rev. {\bf D11} (1975) 3594.




\bibitem{Schechter} J. Schechter, ``Effective Lagrangian With Two
Color Singlet Gluon Fields'', Phys. Rev. {\bf D21} (1980) 3393.

\bibitem{MigdalShifman} A.A. Migdal, M.A. Shifman, ``Dilaton Effective
Lagrangian in Gluodynamics'', Phys. Lett. {\bf 114B} (1982) 445.

\bibitem {zhitnitski} T. Fugleberg, I. Halperin, A. Zhitnitsky, 
``Domain Walls and Theta Dependence in QCD with an Effective
Lagrangian Approach'', hep-ph/9808469.

\bibitem{Aurilia} A. Aurilia, ``The Problem of Confinement: 
From Two to Four Dimensions'', Phys. Lett. {\bf 81B} (1979) 203.

\bibitem{Townsend} A. Aurilia, H. Nicolai, P.K. Townsend,
``Hidden Constants: The $\theta$ Parameter of QCD and the Cosmological
Constant of $N=8$ Supergravity'', Nucl. Phys. {\bf B176}(1980)509.

\bibitem{membrane} G. Gabadadze, ``Modeling the Glueball Spectrum by 
a Closed Bosonic Membrane'',  Phys. Rev. {\bf D58} (1998) 094015; 
hep-ph/9710402. 

\bibitem{VenezianoDi} P. Di Vecchia, G. Veneziano, ``Chiral Dynamics in the
Large $N$ Limit'', Nucl. Phys. {\bf B171} (1980) 253.

\bibitem{Arnowitt}  R. Arnowitt, Pran Nath, ``The $U(1)$ Problem: Current
Algebra and the Theta Vacuum'', Phys. Rev. {\bf D23} (1981) 473; 
``Effective Lagrangians with
$U(1)$ Axial Anomaly. 1, 2'', Nucl. Phys. {\bf B209} (1982) 234, 251;

\bibitem{Trachern} C. Rosenzweig, J. Schechter, C.G. Trahern,
``Is the Effective Lagrangian for QCD a Sigma Model?'', Phys. Rev. 
{\bf D21} (1980) 3388.
 
\bibitem{Pettorino} P. Di Vecchia, F. Nicodemi, R. Pettorino, G. Veneziano,
``Large $N$, Chiral Approach to Pseudoscalar Masses, Mixings and Decays'', 
Nucl. Phys. {\bf B181} (1981) 318.

\bibitem {Pich} A. Pich, ``Chiral Perturbation Theory'', 
Rept. Prog. Phys. {\bf 58} (1995) 563;  hep-ph/9502366.

\bibitem{bubble} M. B. Voloshin, I. Yu. Kobzarev, L. B. Okun',
``Bubbles in Metastable Vacuum'', Yad. Fiz. 
{\bf 20} (1974) 1229 (Sov. J. Nucl. Phys. {\bf 20}
(1975) 644); \\
S. Coleman, ``The Fate of the False Vacuum. 1. Semiclassical Theory'', 
Phys. Rev {\bf D15} (1977) 2929; Erratum-ibid. {\bf D16} (1977)
1248. 

\bibitem{Shifman} M. Shifman, ``Domain Walls and Decay Rate of the Excited 
Vacua in the Large $N$ Yang-Mills Theory'', Phys. Rev {\bf D59} (1999) 021501;
hep-th/9809184.   

\bibitem{bounce} S. Coleman, ``The Fate of the False Vacuum. 1. Semiclassical 
Theory'', 
Phys. Rev {\bf D15} (1977) 2929; Erratum-ibid. {\bf D16} (1977)
1248;\\
C. Callan, S. Coleman, ``The Fate of the False Vacuum. 2. First Quantum 
Corrections'',  Phys. Rev. {\bf D16} (1977) 1762. 

\bibitem{GiaZuraMe} G. Dvali, G. Gabadadze, Z. Kakushadze, 
``BPS Domain Walls in Large $N$ Supersymmetric QCD'', hep-th/9901032. 

\bibitem{orientifold} J. Polchinski, S. Chaudhuri, C.V. Johnson,
``Notes on $D$-branes'', hep-th/9602052; \\
J. Polchinski, ``TASI Lectures on $D$-branes'', hep-th/9611050.

\bibitem{HashimotoOz} A. Hashimoto, Y. Oz, ``Aspects of QCD Dynamics from
String Theory'', hep-th/9809106. 


\end{references}
\end{document}